\documentclass[pra,twocolumn,graphicx,amssymb,floatfix]{revtex4}
\usepackage{graphicx}
\begin{document}

\title{Past of a particle inside an interferometer: the two-state vector approach }

\author{L. Vaidman}
\affiliation{ Raymond and Beverly Sackler School of Physics and Astronomy\\
 Tel-Aviv University, Tel-Aviv 69978, Israel}

\begin{abstract}
   Hashmi {\it et al.} [J. Phys. A  49, 345302 (2016)] claimed that the approach to the past of a quantum particle introduced by Vaidman [Phys. Rev. A 87, 052104  (2013)]  has difficulties in certain examples and that it even can be refuted. Here I reply to their criticism showing that the approach provides a good explanation of  all examples they considered. It is fully consistent with standard quantum mechanics and provides a useful tool for analyzing interference experiments.
       \end{abstract}
\maketitle

\section{Introduction}

 Hashmi {\it et al.}  paper \cite{Hash} is a continuation of the hot discussion about the meaning of the past of a pre- and post-selected particle \cite{past,LiCom,RepLiCom,Danan,morepast,Jordan,Bart,BartCom,Poto,PotoCom}, which has a crucial importance for the controversy regarding counterfactual quantum protocols \cite{Joz,Ho06,Va07,Salih,Gisin,MyCom,SalihReply,Li,count,SalihQT,LiQT,QTCom,LiQTRep}. The authors discussed three setups for which the approach \cite{past} apparently has difficulties, claiming that for their third setup ``the prediction of the theory of the past of the particle is in clear contradiction with standard quantum mechanics''. Standard quantum mechanics does not make any statement about the past of a quantum particle during the time it is inside the interferometer, so there cannot be any contradiction with quantum mechanics. However, in experiments with delicate interference effects there are some subtle effects and it is of interest to discuss internal consistency of the approach.

In the next section I explain the two-state  vector approach to the past of a quantum particle. The three examples of Hashmi {\it et al.} are discussed in Sections 3-5. I claim that they are mistaken in the analysis of the first and the third examples and I disagree with their evaluation of the second example. I conclude the paper in Section 6.

\section{The two-state vector approach to the past of a quantum particle}

The two-state vector approach has its origin in a seminal work of Aharonov, Bergmann and Leibovitz \cite{ABL} in which it was proposed to consider quantum system between measurements in a time-symmetric manner. A  pre- and post-selected quantum system is described by a two-state vector
 \begin{equation}\label{tsv0}
  \langle{\Phi} \vert ~  \vert\Psi\rangle,
\end{equation}
which consists of the usual quantum state  evolving forward in time, $ |\Psi\rangle $, defined by the
results of a complete measurement at the earlier time, and by a quantum state evolving backward in time $ \langle \Phi |  $, defined by the results of a complete measurement at a later time.

  The next important step of the two-state vector formalism (TSVF) was to consider {\it weak measurements} performed on a pre- and post-selected quantum system and discovery that at the weak limit the effective coupling to an observable $O$ is always to the {\it weak value} \cite{AAV,AV90}:
  \begin{equation}\label{wv}
O_w \equiv { \langle{\Phi} \vert O \vert\Psi\rangle \over
\langle{\Phi}\vert{\Psi}\rangle }  .
\end{equation}

The basis of the approach to the past of a pre- and post-selected quantum particle  inside an interferometer \cite{past}  is  the definition:
\begin{quote}
  The particle was present in paths of the interferometer in which it left a weak trace.
\end{quote}

 The motivation for this proposal is that usually we know about presence of objects due to the trace they leave. Associating location of a particle where it left no trace (e.g. Bohmian surrealistic trajectories \cite{Sur,Anal}) does not provide  operational meaning. And since we analyze interferometers, the trace must be weak, otherwise interference would not be observed.

  The crucial issue is the magnitude of the weak trace which is large enough to qualify for the application of the definition. Due to quantum uncertainty, there are tails of quantum waves everywhere. In reality there are no ideal channels in which a particle passing through leaves no trace.
 Thus, a definition according to which the particle is in every region of nonvanishing trace is not acceptable.
 Let us signify $\epsilon$, the parameter quantifying the strength of the interaction in a path of an interferometer: the magnitude of the trace in the path when a single particle passes through. This allows us to make quantitative definition.
  \begin{quote}
  The particle passed through an interferometer was in every  arm in which the trace is of order $\epsilon$.
\end{quote}

This definition of the past of a quantum particle does not rely on the TSVF. Adopting the definition allows  analyzing the past of the particle using standard quantum mechanics. The paradoxical feature of the nested interferometer \cite{Va07,past,Danan} that the particle leaves a disconnected week trace in the inner interferometer is the consequence of standard quantum theory.

 The TSVF provides tools and intuition which simplify the analysis of pre- and post-selected quantum systems:
  \begin{quote}
 The weak trace in the path of  an interferometer is large enough for application of the definition for the past of the particle when the weak value of at least one local observable of the particle in the path does not vanish.
  \end{quote}
   From this follows a simple criterion of the past of the particle:
\begin{quote}
  The particle was present in paths of the interferometer in which there is an overlap of the forward and backward evolving wave functions.
\end{quote}
The overlap is required only in the spatial wave function. Internal degrees of freedom of the particle might be orthogonal, since local operators connecting these degrees of freedom will have nonvanishing weak values and coupling to these operators will lead to a nonvanishing trace. Note, however, that the overlap of the spatial forward and backward evolving wave functions entangled with orthogonal states of an external system will not cause the trace in the path and will not correspond to the presence of the particle in the path.

In the framework of the TSVF we can take these properties as definitions of the past of a quantum particle, but one should not forget that they are rooted, as the whole TSVF, in the standard quantum theory.

\section{Nested interferometer with perfect destructive interference}

The  criteria presented in previous section are criticised by Hashmi {\it et al.} \cite{Hash} by analysis of three examples. The first example is a nested interferometer with perfect destructive interference shown on Fig. 1 (based on Fig. 1 from \cite{Hash}). In Section 1 Hashmi {\it et al.} correctly analyze it in the framework of the TSVF showing the paradoxical feature of presence (according to the weak trace definition) inside the inner interferometer and absence on the way in and the way out of the interferometer. This is represented by a finite value of the weak values of projections on the paths of the inner interferometer  at stage L3:  $({\rm \bf P_C})_w =\frac{t^2}{r^2}$ and $({\rm \bf P_E})_w =-\frac{t^2}{r^2}$ and zero  value of the projections on the paths leading to and out of the inner interferometer  at stages L2 and L4: $({\rm \bf P_B})_w =0$  and    $({\rm \bf P_F})_w =0$, see their Table 1.

\begin{figure}
\begin{center}
 \includegraphics[width=6.2cm]{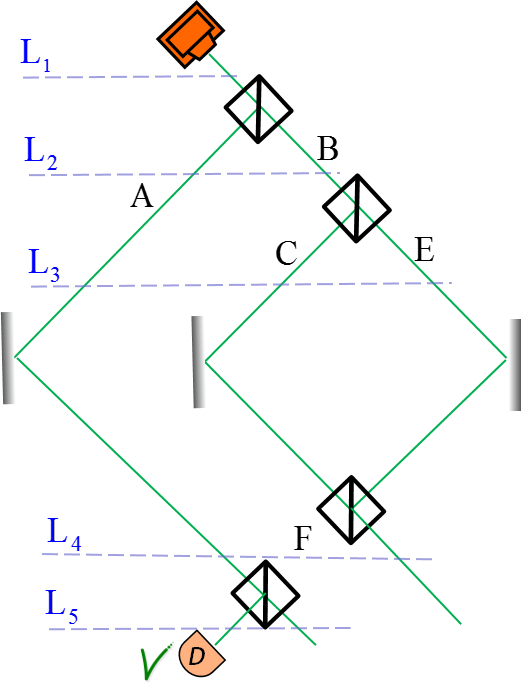} \end{center}
\caption{Nested Mach-Zehnder interferometer.  In the right arm of a large interferometer inserted a small interferometer  tuned such that the particle cannot pass through the right arm of the external interferometer. }
\end{figure}

In Section 2 Hashmi {\it et al.} claim to ``resolve'' the paradox. They introduce weak measurement inside inner interferometer with parameter $\epsilon$ and claim: ``If a weak measurement is performed along the arm F subsequent to the weak measurement inside the inner interferometer then the particle will be revealed along the arm F at the lowest order of the measurement strength parameter.''

I believe, however, that this is not true. Inside the interferometer, the weak measurement reveals the trace of the particle. Hashmi {\it et al.} do not deny this. The size of the signal is given by $\epsilon$, i.e., the first order of the the measurement strength parameter. Thus, the first order is ``the lowest order of the measurement strength parameter''. Assume that the weak measurement in F also has  $\epsilon$ as the measurement strength parameter. The wave leaked to F has the amplitude proportional to $\epsilon$, so the magnitude of the trace  in F will be proportional to $\epsilon^2$. If Hashmi {\it et al.} would complete their Table 2 to include the backward evolving wave function and the weak values as in Table 1, they would see that introduction of weak measurement inside the interferometer leads at stages L2 and L4 to weak values of the projection on the paths towards and out of the inner interferometer $({\rm \bf P_B})_w~=~\mathcal{O}\left(\epsilon\right)$ and $({\rm \bf P_F})_w =\mathcal{O}\left(\epsilon\right)$. This leads to traces of the magnitude proportional to the second order in $\epsilon$ which is postulated to be neglected.

The Hashmi {\it et al.} last comment in Section 2 is correct. The leakage to F is crucial. The results of the experiment \cite{Danan} cannot be explained without it and moreover, the leakage is unavoidable when the trace inside inner interferometer is created. But it does not refute the approach: the  approach  distinguishes different magnitudes of the trace. The trace proportional to the second and higher order of the measurement strength parameter is neglected. It is justified by the fact that at the weak limit of the measurement, such trace is infinitely smaller than the trace defined by the TSVF criterion.

 Although the traces in B and  F are negligible, still there is a conceptual difference between B and F, and places outside the interferometer: placing an opaque object in F  (or B) will change the weak trace inside the inner interferometer, while placing opaque objects outside, will cause no difference. This is why there is a status of ``secondary presence'' of the particle in B and F, see \cite{morepast}.

\section{Nested interferometer with two inner interferometers with perfect destructive interference}

In Section 3 Hashmi {\it et al.} considered another setup with nested interferometers, their Fig. 2. Now, the overlap of the forward and backward evolving wave functions takes place only on a single continues path, so the TSVF predicts that the particle will have a single path. However,  Hashmi {\it et al.} claimed that when weak measurements are performed in two places, this will not be true.

Here their analysis was correct and they mentioned  the resolution of the difficulty: ``One can argue that the trace along the right arm of MZ1 revealed by $P(b1,b2)\propto \epsilon_1 \epsilon_2$ is second order contribution and need to be neglected in the proposed theory \cite{past} based on TSVF.'' But they continued: ``However, this means that TSVF, being a first order theory, should not be applied to the systems with multiple weak measurements, unless the disturbances caused by the weak measurements are properly taken into account.''
Yes, if the weak traces in various paths need to be calculated when some measurements cannot be considered ``weak'', their coupling should  explicitly be taken into account. The two-state vector will be modified and it will provide correct magnitudes of weak traces inside the interferometer on top of the ``weak'' measurements already taken into account.

\section{Hashmi {\it et al.} ``refutation'' of the TSVF approach}

In    Section 4  Hashmi {\it et al.} ``put forward a different argument to refute the theory of the past of the particle \cite{past}''. They modify the first nested interferometer setup by introducing a small phase shift in  arm C of the inner interferometer such that now destructive interference is destroyed. They write: ``In this system the weak measurement can restore destructive interference along a channel that already had a tiny leakage.'' Then the argument goes like this. Since the leakage is crucial for reading the outcome of weak measurement, and the leakage is absent, the weak measurement will not show the presence of the particle inside the inner interferometer, in spite of the fact that the formalism tells us that the particle was there.

The first problem with this argument is the claim that weak measurements can restore destructive interference. Hashmi {\it et al.} mentioned several methods of weak measurements. {\it Any} measurement with external device cannot restore interference. Such weak measurement in arm E  introduces entanglement with external system, so the particle in  arm  E is described by a mixed state which cannot make a complete destructive interference with  a pure state in arm C.

However, if, instead,  we consider a weak measurements in which measuring device is a degree of freedom of the particle passing through the interferometer, as, for exmple, it is the case in the experiment of Danan {\it et al.} \cite{Danan}, then we can consider various unitary evolutions depending on the path of the interferometer, and in particular, the one  canceling  the leakage which is present without the weak measurement operation.

 Placing  a small phase shifter in one arm of the interferometer as Hashmi {\it et al.} provides some information about the path of the particle.
 Indeed, placing such a shifter inside the inner interferometer of the original nested interferometer,  Fig. 1, in arm C (or E) can be considered as a weak measurement of the presence of the particle in this path. Adding another detector $\rm D'$ after the last beam splitter  will allow to observe ``the trace'' inside the inner interferometer  through  the change in the relative intensities  measured in the detectors $\rm D$ and $\rm D'$. This is in contrast to placing the shifter outside the interferometer or,  in arms B and  F where the presence of the particle is not expected, in which case the intensities would not be changed. However, this is not a good method for measurement of the presence of the particle in all paths of the interferometer.  The intensities will not be changed also if the shifter is placed in arm A for which there is a consensus of the presence of the particle. Moreover, this method  does not indicate the presence of the particle passing through a single path. The phase shift does not cause any change in the  intensities at the detectors placed after the beam splitter to which the path leads.

 The idea of Section 4 of Hashmi {\it et al.}  can be better implemented by introducing small transversal shift  of the beam representing weak measurement, instead of the phase shift, conceptually similar to the method of Danan {\it et al.} (but without various frequencies trick). The shift of the beam at the detector will be able to identify the presence of the particle if placed in arms A, C, and E and also the absence of the particle in B and F. Now, the weak measurement in E, i.e., placing the shifter placed in E, will cancel the leakage of the tuned interferometer in which a similar shifter was introduced in C, similarly to the Hashmi {\it et al.} proposal. Here is the difficulty. The shifters do not change  predictions of the TSVF that the particle was present inside inner interferometer, and, in particular, in E. However,  Hashmi {\it et al.} apparently can claim that the weak measurement, represented by placing a  shifter in E  is not be able to reveal the trace of the particle because  there is no leakage out of the inner interferometer. This, however, is not true. Introduction a transversal shifter in E  causes a change in the position of the output beam at the detector. It cancels  the shift introduced by the auxiliary shifter placed in C prior to the weak measurement.
Similarly, in the original experiment proposed in Section 4 of Hashmi {\it et al.}, placing the phase shifter in E restores the  intensities of the detectors $\rm D$ and $\rm D'$ which were changed by phase shifter in C providing a witness of the presence of the particle in E. In this weak measurement experiment the change in the leakage and not the leakage itself is necessary for obtaining information about the presence of the particle in the paths of the interferometer.

\section{Conclusions}
I have shown that  apparent difficulties of the TSVF description of  quantum interference experiments pointed out by Hashmi {\it et al.} are resolved if one performs a more careful analysis. Paradoxical features of traces of pre- and post-selected particles vividly revealed in the framework of the TSVF are fully consistent with standard quantum mechanics. The past of quantum particles in nested interferometers does have these counterintuitive features. The resolution of the   ``paradox'' represented by these surprising  features is not by attempting to show that they are not present,  but  by rejecting classical ``common sense'' arguments which do not hold for quantum particles. Classical way of thinking in terms of continuous trajectories of particles is not supported by the results of quantum experiments with delicate interference effects.

 This work has been supported in part by the  Israel Science Foundation  Grant No. 1311/14  and the German-Israeli Foundation for Scientific Research and Development Grant No. I-1275-303.14.

\end{document}